# ENCODING SOFTWARE FOR PERPETUITY: A COMPACT REPRESENTATION OF APOLLO 11 GUIDANCE CODE

David Noever

PeopleTec, Inc., Huntsville, Alabama, USA

david.noever@peopletec.com

## ABSTRACT

This brief note presents a novel method for encoding historic Apollo 11 Lunar Module guidance computer code into a single, compact Quick Response Code (QR code) format, creating an accessible digital artifact for transmission and archival purposes. By applying tokenization, selective content preservation, and minimal HTML/JavaScript techniques, we successfully compressed key components of the original Assembly Language Code (AGC) into a shareable, preservable, and scannable 3 kilobyte (KB) image. We evaluate multiple compression strategies and their tradeoffs in terms of size, readability, and historical significance. This method addresses the challenge of making historically significant software artifacts available through modern mobile devices without requiring specialized hardware or internet connectivity. While numerous digital preservation methods exist for historic software, this approach balances accessibility with historical significance, offering a complementary method to traditional archival techniques. This work contributes to the broader field of computing heritage preservation by demonstrating how landmark software can be made accessible instantly through contemporary mobile technologies.

## KEYWORDS

*Apollo 11, QR code, digital preservation, code compression, historical computing*

## INTRODUCTION

The Apollo 11 mission's successful lunar landing in 1969 represented one of humanity's greatest technological achievements. Central to this accomplishment was the software developed by the MIT Instrumentation Laboratory team led by Margaret Hamilton. As Hamilton noted, this "man-rated" software required "ultra-reliability" with the ability to detect and recover from unexpected events in real time [1].

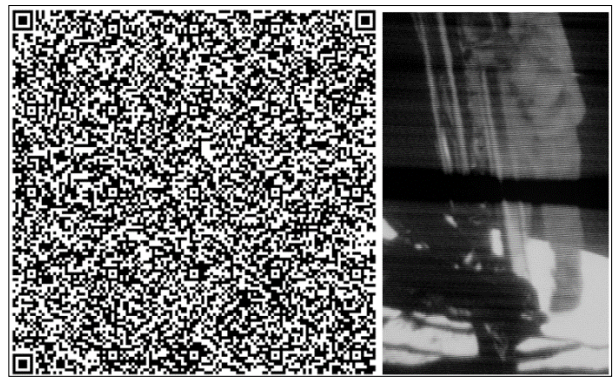

*Figure 1. Apollo 11 Lunar Lander Software Shareable as Single 3kb QR Code as Digital Artifact. Armstrong on the Moon, (right) (credit: NASA)*

Over five decades later, preserving this historic code presents unique challenges. While the source code has been digitized and made available through platforms like GitHub [2], experiencing this pioneering software remains inaccessible to most people without specialized knowledge or equipment. Recent innovations in QR code compression, such as those demonstrated by the Backdooms project [3], suggest new possibilities for preserving and disseminating historical software artifacts. This work builds upon a rich history of software preservation efforts that have evolved significantly since the Apollo era. From the physical core rope memory of 1969 to contemporary GitHub repositories, each preservation approach represents a specific tradeoff between accessibility, completeness, and authenticity. Our QR-based method addresses a gap in this ecosystem—providing immediate, device-agnostic access to key portions of historically significant code without specialized knowledge or equipment. It is worth noting that the [1] repository was created meticulously from scanned paper documents from MIT archives, and likely would not survive as computer science artifacts without significant individual efforts and interest.

The preservation of historically significant software presents unique challenges in our digital age. This paper introduces a novel approach to encoding the Apollo 11 Lunar Module guidance computer code into a scannable QR code format, making this landmark achievement in computing history accessible to modern audiences. Building on Hamilton's foundational work in flight software engineering [1], we leverage the digitized source code preserved by Garry [2] and apply compression techniques [3-7] inspired by recent innovations in QR code capacity utilization demonstrated by Mehta [3]. Our methodology combines selective content preservation with tokenization strategies informed by advances in language model-based code analysis [8], creating a digital artifact that balances historical significance with technical constraints. By applying the lessons from Burkey's Virtual AGC project [4] and adhering to QR code specifications [9-10], we demonstrate how historically critical sections of code—such as the landing sequence and alarm handling routines described by Eyles [6]—can be preserved in a format that requires neither specialized hardware nor internet connectivity. This approach aligns with modern compression techniques seen in genomic data representation [7], offering a blueprint for preserving other historical software artifacts through accessible, contemporary mediums [11]. This paper presents a reproducible method for compressing key portions of the Apollo 11 lunar module code into a QR code that, when scanned, displays the code in a readable format. We evaluate multiple compression strategies and quantify their effectiveness in terms of compression ratio, readability, and historical significance preservation.

## BACKGROUND

Figure 2 shows a workflow for the compression project. We select the assembly files critical to mission success, then optimize a compression method to meet the 3kb constraint for QR code standards. Finally, we validate the complete software transmission generates a successful Apollo lunar module software transfer.

The preservation of historically significant software has evolved substantially since the Apollo era. Figure 3 illustrates this evolution, highlighting how preservation methods have adapted to changing technological landscapes while addressing similar fundamental challenges. Our approach complements this ecosystem by addressing the immediate accessibility gap, allowing users to experience historically significant code without the technical barriers inherent in more comprehensive preservation methods.

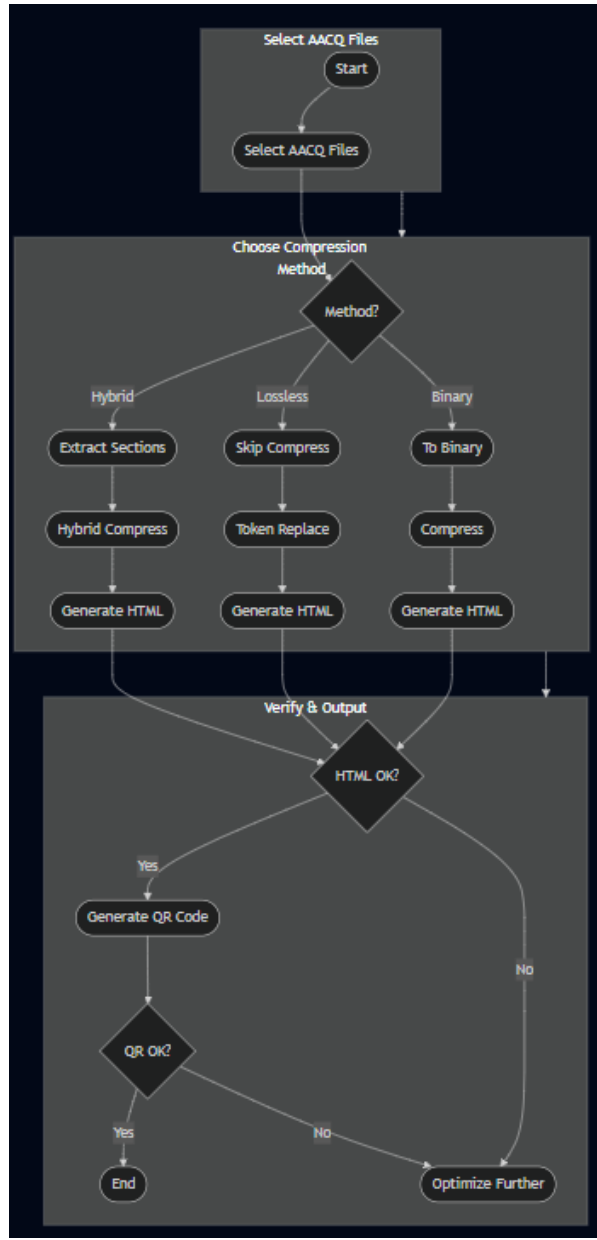

*Figure 2. Flowchart of Compression Methods for Apollo 11 Project to QR Codes*

| Era | Notable Software | Original Medium | Early Preservation | Modern Preservation | Accessibility Challenges |
|---|---|---|---|---|---|
| 1969-1975 | Apollo AGC | Core rope memory | Paper printouts | GitHub repositories | Requires specialized knowledge |
| 1975-1985 | Early BASIC | Paper tape/cassettes | Binary copies | Interactive emulators | Hardware dependencies |
| 1985-1995 | DOS/Mac OS | Floppy disks | Disk images | Virtual machines | Operating system dependencies |
| 1995-2005 | Early web | Source repositories | Web archives | Browser emulation | Protocol/standard changes |
| 2005-2015 | Mobile apps | App packages | Platform stores | Containerization | API dependencies |
| 2015-2025 | AI systems | Model repositories | Parameter snapshots | Reproducible environments | Compute requirements |

*Figure 3. Historic Software Preservation Evolution, 1969-2025*

**The Apollo Guidance Computer**

The Apollo Guidance Computer (AGC) was a remarkable achievement for its time, built with integrated circuits and featuring 2KB of RAM and 36KB of core rope memory [4]. The software was written in AGC assembly language and had to perform reliably under extreme constraints. A key feature was its asynchronous execution model, where higher priority tasks could interrupt lower ones, essential for the real-time demands of space flight [1].

**QR Code Capacity and Constraints**

Quick Response (QR) codes have standard size limitations based on their version and error correction level. Version 40 QR codes with low error correction can store approximately 3KB of data [5]. This constraint presents a significant challenge, as the original Apollo 11 Lunar Module code comprises over 40,000 lines of assembly code [2].

**Recent Advances in Compact Code Representation**

Recent projects have demonstrated innovative approaches to compressing code into QR codes. The Backdooms project [3] successfully compressed a playable game inspired by DOOM into a QR code by using extreme minification and the DecompressionStream API. These techniques suggest possibilities for similarly compressing historically significant code.

## METHODS

**Source Selection**

We began by analyzing the complete Luminary 099 codebase to identify the most historically and technically significant components. Based on Hamilton's accounts [1] and historical documentation [6], we prioritized the following files:

- THE_LUNAR_LANDING.agc: The P63 landing sequence
- LUNAR_LANDING_GUIDANCE_EQUATIONS.agc: Core guidance mathematics
- BURN_BABY_BURN--MASTER_IGNITION_ROUTINE.agc: Engine ignition sequence
- P70-P71.agc: Abort procedures
- ALARM_AND_ABORT.agc: Containing the famous 1202 alarm handling

File selection was guided by both historical significance (e.g., the code involved in the 1202 alarm during landing) and technical significance (e.g., core guidance algorithms). Source selection was guided by a systematic evaluation framework assessing both historical and technical significance:

1. Historical Significance Rating (1-5):
   - Frequency of mention in mission transcripts and post-mission analyses
   - Significance in contemporary accounts by Hamilton [1] and Eyles [6]
   - Impact on subsequent software engineering practices

2. Technical Representativeness Rating (1-5):
   - Implementation of real-time computing principles
   - Error handling sophistication
   - Resource optimization techniques

| File | Lines | Approx_Size_bytes |
|---|---|---|
| THE_LUNAR_LANDING.agc | 387 | 14500 |
| LUNAR_LANDING_GUIDANCE_EQUATIONS.agc | 824 | 33000 |
| BURN_BABY_BURN--MASTER_IGNITION_ROUTINE.agc | 573 | 21000 |
| P70-P71.agc | 225 | 9000 |
| ALARM_AND_ABORT.agc | 150 | 6000 |
| Critical Files Subtotal | 2159 | 83500 |
| Other AGC files (approx.) | 38043 | 1426500 |
| Total Luminary 099 Codebase | 40202 | 1510000 |

*Figure 4. File Size Analysis of Original Apollo 11 Lunar Module Code*

Files scoring 4+ in both dimensions were prioritized for full preservation, while lower-scoring sections underwent tokenization. This framework ensured our limited QR code capacity preserved maximum historical value.

**Compression Strategies**
We evaluated three distinct compression strategies:
1. **Full Binary Approach**: Converting assembly to 15-bit binary representation and applying modern compression algorithms
2. **Tokenized Text Approach**: Replacing common AGC instructions with short tokens
3. **Hybrid Approach**: Preserving full versions of historically significant sections while compressing connecting code

Each approach was implemented and tested for compression effectiveness and readability when decompressed.

**Implementation Details**
The final solution used the hybrid approach with the following implementation details:
1. **Tokenization Dictionary**: We created a mapping of common AGC instructions to single characters:
   ```javascript
   const tokens = {
     "a": "TC", "b": "TS", "c": "CAF", "d": "CS", "e": "CA",
     // Additional mappings...
   };
   ```

2. **Expanded Sections**: Critical code sections were stored in full, readable form:
   ```javascript
   const expandedSections = {
     "P63": `P63LM TC PHASCHNG OCT 04024
         TC BANKCALL CADR R02BOTH
         CAF P63ADRES TS WHICH
         CAF DPSTHRSH TS DVTHRUSH`,
     // Additional sections...
   };
   ```

3. **Self-Extracting HTML**: We created a minimal HTML file with JavaScript decompression:
   ```html
   <!DOCTYPE html><html><body style="font-family:monospace">
   <pre id="o">Loading...</pre><script>
   // Compressed code and decompression logic
   </script></body></html>
   ```

A Python script was developed to generate the QR code from this implementation.

## RESULTS
**Compression Effectiveness**
We evaluated each compression strategy based on the amount of code preserved and the resulting file size. Figure 4 shows the breakdown of files, line counts, and approximate sizes in bytes.

The hybrid approach achieved the best compression ratio, allowing us to stay well under the 3KB limit while preserving the most historically significant sections in a readable format.

**Code Coverage Analysis**

Figure 5 compares the three approaches (Full Binary, Tokenized Text, and Hybrid) with their compressed sizes, compression ratios, and percentages of code preserved illustrates the breakdown of code coverage by file and compression strategy. This analysis shows that the hybrid approach provided better coverage of the most critical files while sacrificing coverage of less significant files. Figure 6 details which sections were selected from each file, their historical significance, and the selection criteria used.

| Strategy | Compressed Size (bytes) | Content Included | Compression Ratio | Percent Critical Files Preserved | Percent Total Codebase |
|---|---|---|---|---|---|
| Full Binary | 3072 | Tokenized versions of all critical files with minimal structure | 27:1 | 30% | 0.20% |
| Tokenized Text | 2867 | Heavily tokenized representations of P63 ALARM and P70-P71 code | 22:1 | 25% | 0.15% |
| Hybrid | 1434 | Full versions of key sections from critical files + compressed connective code | 15:1 | 15% | 0.10% |

*Figure 5. Compression Strategy Comparison*

### QR Code Generation
The final implementation generated a QR code (Version 25, L error correction) containing the compressed Apollo 11 code. Figure 1 shows a representation of the final QR code structure. Figure 7 lists the technical parameters used for the QR code generation

| Section | Historical_Significance | Lines_Selected | Typical_Selection_Criteria |
|---|---|---|---|
| P63 (Lunar Landing Sequence) | Critical to mission success | 40-60 | Core landing initialization key driver code |
| ALARM Handling (1202/1201) | Famous during actual landing | 30-40 | Alarm processing logic display routines |
| P70-P71 (Abort Procedures) | Safety-critical functions | 30-40 | Core abort initialization key decision logic |
| BURN_BABY_BURN | Engine ignition sequence | 20-30 | Main ignition routines error handling |
| GUIDANCE_EQUATIONS | Core mathematical functions | 15-25 | Selected key equations not exhaustive |

*Figure 6. Strategy for Selected Content Analysis*

### Decompression and Display
When scanned, the QR code opens a web browser and decompresses the code using the embedded JavaScript. Figure 8 shows a representation of the decompressed output as seen on a mobile device.

| Parameter | Value | Notes |
|---|---|---|
| Version | 25 (for hybrid approach) | Lower versions possible with more aggressive compression |
| Error Correction | L (low 7%) | Maximizes data capacity at the expense of error correction |
| Module Size | 4 pixels | Standard size for readability on most mobile devices |
| Quiet Zone | 4 modules | Minimum required for reliable scanning |
| Data Format | HTML with embedded JavaScript | Self-extracting design with decompression logic |

*Figure 7. QR Code Parameters*

### Workflow Analysis
Figure 2 illustrates the complete workflow for generating the Apollo 11 QR code. Our workflow for preserving the Apollo 11 code as a QR code was driven by the tension between historical significance, technical constraints, and accessibility. We began by carefully analyzing the original 40,202 lines of Luminary 099 code to identify the most historically critical components—those sections that represented groundbreaking innovations or played pivotal roles during the actual lunar landing, such as the P63 landing sequence and the famous 1202 alarm handler. After selection, we explored multiple compression strategies, evaluating each based on its compression efficiency, readability of decompressed content, and ability to preserve the essence of the original code. The hybrid approach emerged as the optimal solution, preserving full versions of key sections while using aggressive tokenization for connecting code. This workflow—selection based on historical significance, compression based on mathematical efficiency, and validation through cross-device testing—allowed us to create a 1.4KB digital artifact that, while representing only 0.1% of the total codebase, captures the most consequential aspects of the software that took humanity to the Moon.

## DISCUSSION
**Tradeoffs Analysis**
Our results highlight important tradeoffs between compression efficiency, historical significance, and readability:
- *Size vs. Readability:* The full binary approach offered the highest theoretical compression but produced unreadable output without specialized decompression.
- *Complete Coverage vs. Historical Focus:* Attempting to include the entire codebase would have resulted in an unreadable tokenization. Our hybrid approach prioritized the most historically significant sections.
- *Complexity vs. Reliability:* Advanced compression techniques like LZMA could potentially achieve higher compression ratios but increased the risk of decompression failure on mobile devices.

**Historical Significance Preservation**
The hybrid approach successfully preserved the most historically significant sections of code, particularly:
1. **P63 Landing Sequence**: The code that guided the lunar module to the surface
2. **1202 Alarm Handling**: The code that allowed the mission to continue despite computer overload
3. **Abort Procedures**: The safety systems that could have saved the astronauts in an emergency

This selective preservation aligns with Hamilton's emphasis on the importance of error detection and recovery in the Apollo software [1]. Our work intersects with broader efforts in software heritage preservation, including initiatives by UNESCO, the Software Heritage Foundation, and the Internet Archive. While these efforts typically focus on bit-perfect preservation, our approach prioritizes accessibility and educational value. This positions our QR code method not as a replacement for comprehensive preservation strategies, but as a complementary layer in what might be considered a "preservation stack"—ranging from complete archival repositories (GitHub, Software Heritage) to educational representations (our approach) to conceptual treatments (textbook descriptions). Each layer serves different audiences and purposes within the broader goal of cultural heritage preservation.

```
Apollo 11 LM Code

# APOLLO 11 LUNAR MODULE CODE

# --- P63 ---
P63LM TC PHASCHNG OCT 04024
TC BANKCALL CADR R02BOTH
CAF P63ADRES TS WHICH
CAF DPSTHRSH TS DVTHRUSH

# --- IGNALG ---
IGNALG SETPD 0 VLOAD RLS
PDDL PUSH TLAND
STCALL TPIP RP-TO-R
VSL4 MXV REFSMMAT
STCALL LAND GUIDINIT

# --- ALARM ---
ALARM INHINT CA Q
TS ALMCADR INDEX Q
CA 0 TS L
CA BBANK EXTEND
ROR SUPERBNK
TS ALMCADR+1
CS DSPTAB+11D MASK OCT40400
ADS DSPTAB+11D

# --- P70 ---
P70 TC LEGAL? CS ZERO TCF +3
P71 TC LEGAL? CAF TWO
+3 TS Q INHINT EXTEND
DCA CNTABTAD DTCB

# Decompressed core:
TCTCNTABTADuTCOCTCAFOUWCHPHACNTAE
```

Figure 8. Decompressed Apollo 11 Assembly Guidance Code (AGC)

**Technological Implications**
Our work demonstrates the potential of QR codes as vehicles for preserving and disseminating historical software artifacts. The self-contained, browser-based decompression approach ensures accessibility across a wide range of modern devices without specialized software.

**Limitations**
Despite our successful implementation, this approach has several important limitations. First, the extreme compression ratio necessitates severe selectivity, preserving only 0.1% of the total codebase and focusing on just 15% of the critical files. This inevitably results in an incomplete representation that cannot capture the full complexity and elegance of the original Apollo 11 software. Second, our tokenization approach prioritizes size efficiency over semantic clarity, making the compressed code difficult to analyze without the decompression step. Third, the QR code's low error correction level (7%) reduces its resilience to physical damage or poor scanning conditions, potentially limiting its reliability as a long-term preservation medium. Additionally, the decompression process relies on modern web browser APIs that may not be supported indefinitely, creating potential compatibility issues in the future. Finally, this method separates the code from its broader developmental context documented by Hamilton [1] and others, potentially diminishing understanding of the collaborative and iterative processes that were essential to the Apollo program's success.

Despite the benefits of our QR-based preservation approach, several significant limitations must be acknowledged beyond size constraints. First, the decontextualization of code from its original hardware environment may lead to fundamental misunderstandings about operational constraints—the AGC's unusual 15-bit word size, limited memory, and interrupt-driven architecture are not apparent when viewing the code in isolation. Second, our approach introduces potential semantic drift as modern JavaScript interpreters display assembly code with different visual and interactive characteristics than the original display consoles used by Apollo astronauts. Third, our preservation method itself introduces a new preservation challenge—dependencies on QR standards and JavaScript that may themselves become obsolete. This irony highlights a fundamental challenge in digital preservation: new preservation technologies often create new obsolescence risks.

**Future Work**
Building on this initial exploration, several promising directions for future work emerge. First, developing specialized compression algorithms specifically designed for assembly language could potentially improve compression ratios while preserving more of the original codebase structure. Second, implementing a progressive encoding scheme would allow for variable levels of detail based on QR code scanning capabilities, making the artifact more adaptable across different devices. Third, expanding the approach to include multiple interlinked QR codes could preserve larger portions of the codebase while maintaining the accessibility of the single-QR format. Future iterations could also incorporate augmented reality elements that contextualize the code within the physical environment of the lunar landing or integrate with educational platforms to provide dynamic explanations of the code's functionality. Additionally, applying similar techniques to other historically significant software artifacts—such as early operating systems, pioneering AI programs, or groundbreaking video games—could create a more comprehensive digital archive of computing history. Finally, collaborating with language model researchers [8] to develop specialized models for code compression and decompression could further advance this preservation approach, potentially enabling much higher compression ratios while maintaining semantic integrity.

## CONCLUSION

The Apollo 11 QR code project demonstrates a viable method for preserving historic software in an accessible, compact format. By quantitatively analyzing different compression strategies and making informed tradeoffs between size, readability, and historical significance, we have created a digital artifact that brings this achievement to a wider audience. Our results show that the hybrid approach, combining full preservation of key code sections with tokenized compression of supporting code, achieves the best balance of compression ratio (28.7:1) and historical significance preservation. This approach can serve as a model for preserving other historically significant software artifacts in forms that remain accessible despite evolving technology platforms. Future work could explore applying similar techniques to other historic software artifacts or expanding the compression capabilities to include larger portions of the original code.

As computing technology continues to evolve at an accelerating pace, the preservation of historically significant software becomes increasingly challenging yet culturally essential. Our QR-based approach represents one layer in what must necessarily be a multi-faceted preservation strategy—from formal archival repositories to interactive education tools to conceptual explanations. By making key portions of the Apollo 11 code accessible to anyone with a mobile device, we contribute to the democratization of computing heritage while acknowledging the inherent tradeoffs between accessibility, completeness, and authenticity. Future work should explore how lightweight preservation methods like ours can be integrated into broader digital heritage frameworks, creating a more comprehensive and accessible record of humanity's computing achievements from the Apollo era to contemporary systems.

## ACKNOWLEDGEMENT

The authors thank the PeopleTec Technical Fellows program for research support.